\documentclass[%
 reprint,
superscriptaddress,
 amsmath,amssymb,
prl
]{revtex4-1}

\usepackage{graphicx}
\usepackage{subfigure}
\usepackage{dcolumn}
\usepackage{bm}
\usepackage{color}%
\usepackage{amsmath}
\usepackage{verbatim}
\usepackage{mathtools}
\usepackage{epstopdf}
\usepackage{float}

\def\beq{\begin{equation}}
\def\eeq{\end{equation}}

\begin{document}


\title{Collision-driven emergence of the cosmic web}

   \author{Raphael Blumenfeld}
   \email{rbb11@cam.ac.uk}
\affiliation{Gonville \& Caius College, Cambridge University, Trinity St., Cambridge CB2 1TA, UK}

\date{\today}

\begin{abstract}
Gravitational-collapse-based explanations of the cosmic web lead to problems in estimating the total mass in the universe.
A first-principles several-scales model is developed here for the structural organisation of cosmic matter in a flat universe, showing that the web formation could be driven by inelastic collisions before gravity took hold, suggesting a possible way to resolve these problems. 
The following results are derived. 
(i) The diffusion rate in the particulate gas after recombination is sub-anomalous, with a rapid decay of particle velocities. 
(ii) The evolution of the particle velocity distribution is calculated explicitly.
(iii) The gas density is shown to be unstable, leading to void formation and clusters nucleation. 
(iv) Rounded clusters are shown to be unstable and tend to elongate. 
(v) An equation is derived for the growth of long clusters into filaments and solved explicitly. 
The fast-growing clusters deplete the regions around them and generate large voids, potentially giving rise to the cosmic web before gravity dominated. 

\end{abstract}



\maketitle

\noindent{\bf Introduction}

The large scale structure of galaxies and dark matter comprises filaments and sheets, known as the cosmic web \cite{Spetal06,Aretal07}. A number of models, including cold dark matter \cite{vdWB08, Caetal14,CDM}, have been proposed for the formation of this structure, mainly based on gravity as the mechanism to destabilise the primordial uniform density into clusters. 
The aim here is to show that even a minute energy dissipation in particle collisions could lead to such a structure before gravity could take hold. This relieves gravitational collapse from carrying the full responsibility for the formation of the cosmic web, potentially alleviating the problem in estimating the universe's total mass. 
The proposed model starts at recombination and progresses through several length and time scales. 
The following analysis is based, for simplicity, on several assumptions:
(i) the universe is flat;
(ii) after recombination, all matter consists of charge-neutral non-relativistic particles in a gaseous state of relatively uniform density with a mean free path much larger than particle sizes;
(iii) particles do not exchange charge on collisions and electrostatic and dipolar interactions are negligible;
(iv) particles interact negligibly with radiation. 
For clarity, the universe's expansion is neglected in some of the calculations, but this does not affect qualitatively the main conclusions.

Gravity-driven clustering encounters what is known as the bouncing barrier problem~\cite{BouncBar}. For gravity to be relevant for two similar particles of masses $m$ and sizes $D$, their kinetic energies must satisfy
\begin{equation}
\frac{G m}{D} \geq \frac{v_r^2}{2} \ .
\label{Gravity}
\end{equation}
For example, using a value of $G$ similar to today's \cite{G}, this requires that the relative velocity of two neutral hydrogen be below $10^{-14}$m/s. But with such low velocities they are already in a cluster. What, then, could lead to such proximity in the first place? 
The perturbative cosmological structure formation framework~\cite{MaBe95} ameliorates the bouncing barrier problem, but does not eliminate it completely. 
Nor does the idea that pre-recombination dark matter started collapsing gravitationally well before recombination, as evidenced by the problems in estimating the total universe's mass. Velocities attenuation because of the universe's expansion also do not lead to such low kinetic energies over timescales that are relevant to the age of the universe. 
It is claimed here that a slight collisional inelasticity of the neutral post-recombination particles is cosmologically relevant to achieve clustering. 

Models in the literature commonly use either continuum fluid-dynamics or thermodynamic equations of state \cite{Pe80,Uhetal18}. Both these approaches are problematic for inelastically colliding particles. Conventional flow equations have been shown to fail to capture accurately the rheology of dense such fluids \cite{MiDi,FoPo08,ScBl11,ScBl18} and equations of state, based on equilibrium thermodynamics also fail for far-from-equilibrium systems \cite{Edwards89a,Edwards89b,BlEd03,BlEd06}. The modelling approach presented here circumvents these problems.  
It is shown here that inelastic collision dynamics not only give rise to clustering but also explain the ubiquity of filaments and sheets. The modelling focuses on neutral hydrogen, but could be extended to explain the abundance of the dark matter in the cosmic web, as discussed in the concluding section. 

After recombination, matter could be regarded as a dense gas of classical particles moving in a flat universe at velocities many orders of magnitude faster than $v_r$, about $\cal{O}$($10^3-10^4$)m/s. The model is constructed as follows.
Firstly, the diffusion in such a gas is shown to be `sub-anomalous' in that a particle's mean square distance (MSD) grows {\it logarithmically} with time. 
Secondly, using this result, the evolution of the particle velocity distribution is derived explicitly and is shown to decay rapidly, even when the universe's expansion is neglected. 
Thirdly, using both the above results, the uniform density is shown to be {\it unstable}, with less dense regions `bleeding' particles into denser ones. This gives rise to nucleation of clusters. 
Fourthly, clusters are shown to grow by `capturing' colliding particles with insufficient momenta to escape and an equation for the growth rate of the nuclear clusters is constructed and solved explicitly. The growth rate is shown to be shape-dependent with filamentary and sheet-like clusters growing faster than compact ones. 
While some of these mechanisms have been discussed individually in the literature \cite{Lietal17}, their integration into a coherent picture and the new results derived here give this model a predictive power, making it a suitable jumping board for better insight into the organisation of the cosmic structure. The results are consistent with existing numerical observations and their ramifications are discussed. \\

\noindent{\bf Diffusion in a cooling gas}

Starting from the particle scale, consider a general particulate gas in three dimensions with relatively high kinetic energes. For simplicity, all particles are assumed to be similar. The particles move ballistically between collisions, on each of which they lose roughly a constant small fraction, $\epsilon$, of their momenta. The loss can be either by radiation or, on slow collisions with loose groups of particles, by converting their kinetic energy to vibrations in the group on impact. A weak dependence of $\epsilon$ on velocity, as observed in \cite{epsv1,epsv2,Gretal09}, can be included without affecting qualitatively the general conclusions. 
The small drag between collisions in an expanding universe can also be included, but this would affect little the following results. The main effect of the drag would be to slow down somewhat the processes described below.  
Assuming an isotropic velocity distribution and an initially uniform particle number distribution, $\rho$, within a region of space, the statistics of the particle trajectories are as in conventional random walks. The only difference is that the momentum loss stretches the time spent between two successive collisions by $1/\epsilon$. With a mean free path $l_0=\rho^{-1/d}$ ($d=2,3$), the MSD of such a random walk after $N$ steps is $\langle R^2(N)\rangle = dl_0^2 N$ and the time to make $N$ steps, starting at velocity $v_0$, is 
\begin{equation}
t = \frac{l_0}{v_0}\sum_{k=1}^N \epsilon^{1-k} = \frac{\epsilon^{1-N}-\epsilon}{1-\epsilon}\ \frac{l_0}{v_0} \ .
\label{Time}
\end{equation}
Inverting (\ref{Time}), the velocity after the $N$th step is
\begin{equation}
v_N=v_0\epsilon^N= \frac{v_0}{1+t/\tau_0} \ ,
\label{Speed}
\end{equation}
with $\tau_0\equiv \epsilon l_0/[(1-\epsilon) v_0]$. The MSD is then
\begin{equation}
\langle R^2\rangle = \frac{\ln{\left(1+t/\tau_0\right)}}{\ln{(1/\epsilon)}}dl_0^2  \ .
\label{MSD1}
\end{equation}
From (\ref{Time}) and (\ref{Speed}), one obtains that the kinetic energy decays as $1/\left(1+t/\tau_0\right)^2$. This decay resembles Haff's law \cite{Haff}, derived for particulate gases that dissipate energy differently - by viscosity \cite{BrPo00}. The current derivation through random-walk formalism has an advantage over standard derivations of Haff's law that use a combination of continuum flow equations and equations of state, both of which involve additional assumptions about the particulate medium. Moreover, this derivation provides the explicit dependence on the readily estimated quantities $v_0$, $l_0$ and $\epsilon$. 
The derived logarithmic increase in (\ref{MSD1}) is `sub-anomalous' - it is slower than in anomalous diffusion, $\langle R^2\rangle\sim t^{\alpha(<1)}$. \\

\noindent{\bf The velocity distribution}

\begin{figure}[htbp]
   \centering
   \includegraphics[width=.4\textwidth]{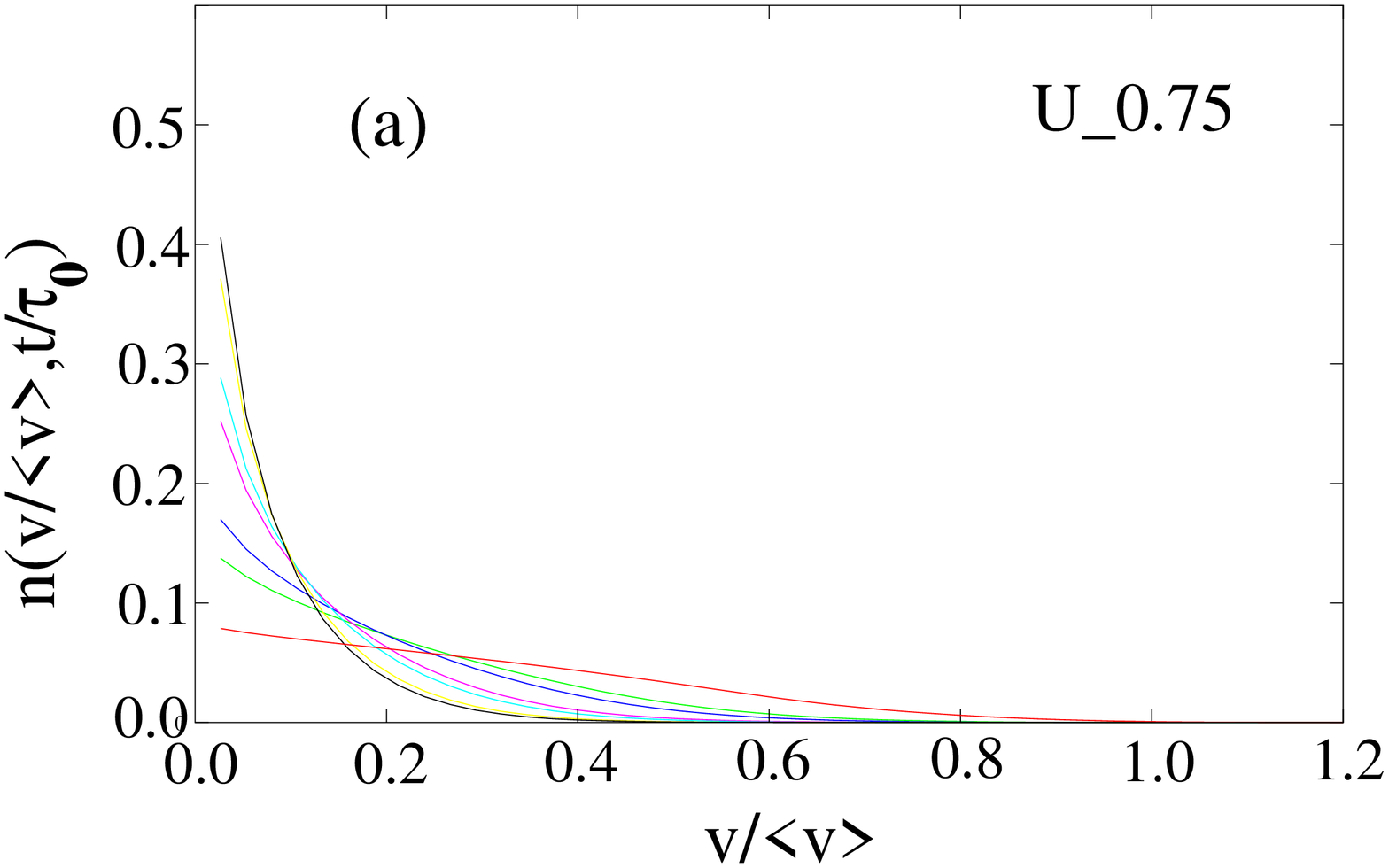}
   \includegraphics[width=.4\textwidth]{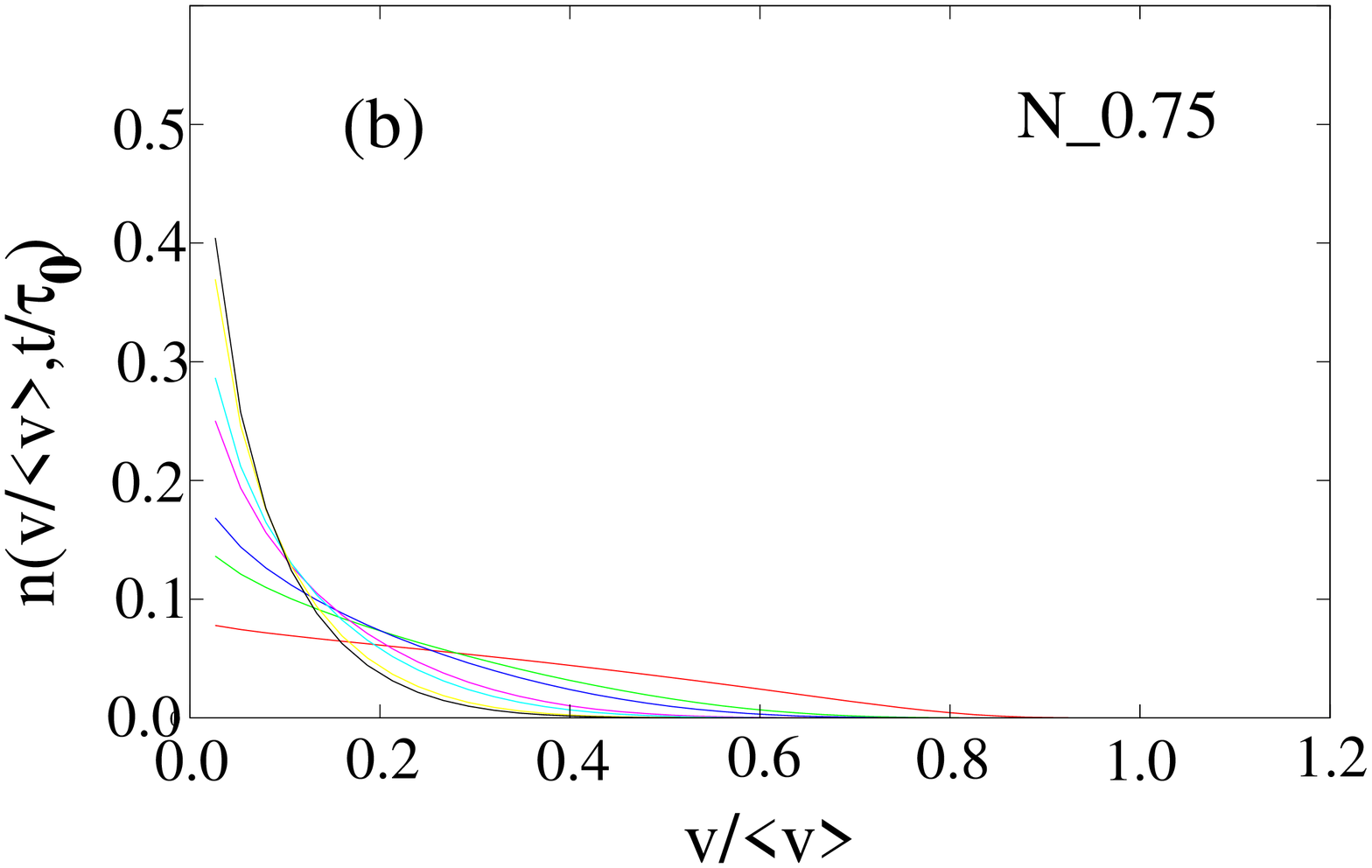}
   \caption{A generic evolution of the unconditional velocity PDF for an initial uniform PDF (a) and normal PDF (b), shown at times: $t = 0.1$ (red), $0.5$ (green), $1.0$ (blue), $5.0$ (purple), $10.0$ (light blue), $50.0$ (yellow), $100.0$ (black). The inter-particle restitution coefficient is $0.75$. The differences between these two cases in the convergence to a delta-function as $t\to\infty$, as well as compared with restitution coefficients $0.50$ and $0.90$, are hardly noticeable.}
   \label{fig:Pv}
\end{figure}

Assuming isotropic uniform density and velocity distributions at some initial time, $t=0$, the probability that a particle, with speed $v_0$, covers a distance $R$ and experiences $N$ collisions is $P_N(N,R)=Ae^{-dR^2/Nl_0^2}$, with $A$ a normalisation factor. 
After $N$ collisions, and before any cluster nucleates, the particle's speed is $v_N=v_0\epsilon^N$. Since $N$ is distributed, the probability that $v_N$ is between $v$ and $v+dv$ is 
\begin{equation}
p(v,t\!\mid\! v_0) dv = \sum_{N=1}^\infty\! P_N(N,R)\ \delta\!\left( N-\frac{\ln{\left(v_0/v\right)}}{\ln{\left( 1/\epsilon\right)}}\right)\! dv 
\label{PDFv1}
\end{equation}
and substituting for $P_N$ yields for this particle
\begin{equation}
n(v,t\!\mid\! v_0) \equiv \rho p(v,t\!\mid\! v_0) = C \rho\left(1+t/\tau_0\right)^{1/\ln{\left(v/v_0\right)}} ,
\label{PDFv2}
\end{equation}
with $0<v< v_0$. The normalisation factor is
\begin{equation}
C=C(v_0,t) = \frac{1}{2v_0\sqrt{\ln{\left(1+t/\tau_0\right)}}\mid\! K_1\left(2\sqrt{\ln{\left(1+t/\tau_0\right)}}\right)\!\mid} \ ,
\label{NormC}
\end{equation}
where $K_\nu(x)$ the modified Bessel function of the second kind \cite{GrRy}. 
Given an initial speed distribution $n_0\left(v_0\right)$, the distribution at later time is then
\begin{equation}
n(v,t) = \int\limits_{v}^{v_{max}} C(v_0,t) \left(1+t/\tau_0\right)^{1/\ln{\left(v/v_0\right)}} n_0(v_0) dv_0 \ ,
\label{PDFv3}
\end{equation}
with $v_{max}<\infty$ the highest possible speed at $t=0$. This integral is difficult to calculate analytically. Evaluating it numerically for a uniform and normal forms of $n_0\left(v_0\right)$, with $\epsilon=0.50, 0.75$, and $0.90$, shows convergence with time to very similar distributions. Typical examples are shown in Fig. \ref{fig:Pv} for $\langle v_0\rangle=30$, $\epsilon=0.75$ and $\rho = 1$ (arbitrary units). 
Since relation (\ref{MSD1}) and the distribution of $N$ are valid for sufficiently many collisions only the distributions for $t/\tau_0 \gtrapprox {\cal{O}}(10)$ are practically relevant. 
The distribution (\ref{PDFv3}) can be approximated, using the mean value theorem:
\begin{equation}
n(v,t) = C(u_0,t) \rho \left[1+ \frac{(1-\epsilon)u_0}{l_0}t\right]^{1/\ln{\left(v/u_0\right)}}  \ ,
\label{PDFv4}
\end{equation}
with $0<v<u_0<v_{max}$. This expression is exact for $n_0(v_0)=\rho\delta\left(v_0-u_0\right)$. \\

\noindent{\bf Unstable density fluctuations} 

\begin{figure}[htbp]
      \centering
      \includegraphics[width=.4\textwidth]{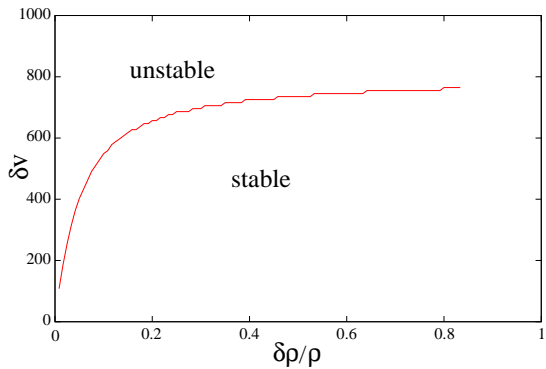}
      \caption{A typical density stability diagram for $\epsilon = 0.8$, $\rho_1=1./l_1^3$ and $v_1 = 1000. l_1/\Delta t$.}
      \label{fig:StabilityPhases}
   \end{figure}

There is significant separation of time scales between the diffusion process and large-scale density changes. Eq. (\ref{MSD1}) describes the diffusion of one particle within a relatively homogeneous region and is valid after a sufficient number of collisions, i.e. $t \geq \tau_d \gg l_0/(\phi v_0)$. Density changes require many particles to `diffuse' out of or into a region, a process that takes place over times much longer than $\tau_{d}$. 
Suppose a local density fluctuation in some region, $A$, reducing $\rho_A$ relative to $\rho_B$ in a neighbour region $B$. 
The flux of particles between the regions depends on the density and velocities and, although $\rho_B>\rho_A$, the higher rate of collisions in $B$ reduces the velocities faster than in $A$. 
   \begin{figure}[htbp]
      \centering
      \includegraphics[width=.15\textwidth]{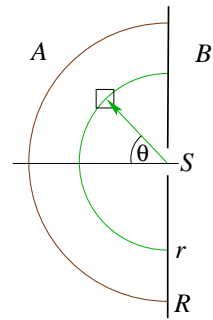}
      \caption{Only particles of speed $v$, which are closer than $R(\Delta t)=\sqrt{\ln{\left(1+(1-\epsilon)v\Delta t/\left(\epsilon l_0\right)\right)}/\ln{(1/\epsilon)}}l_0$ to the small area $S$, contribute to the flux through $S$.}
      \label{fig:FluxFlat}
   \end{figure}
Assuming an isotropic velocity distributions in both regions, consider the number of $A$ particles hitting a small area $\delta S$ in the boundary between $A$ and $B$ within a short time interval $\Delta t$.
Fixing the origin at $\delta S$, the probability that an $A$ particle with speed $v$ starting from the infinitesimal volume element $dV=r^2sin\theta d\theta d\phi$ at $(r,\theta,\phi)$ (Fig. \ref{fig:FluxFlat}) does not hit the boundary before reaching $\delta S$ is $(1+\cos{\theta})/2$. Multiplying by the relative solid angle extended by $\delta S$, its probability to arrive at $\delta S$ is 
\beq
p_s = \frac{1+\cos{\theta}}{2}\frac{\delta S\cos{\theta}}{4\pi r^2} \quad .
\label{ps}
\eeq
The number of $A$ particles originating from $dV$ and hitting $\delta S$ during $\Delta t$ is then $d\Delta{\cal{N}}_A =   p_s n_A(v) dV dv$. Integrating over $0\leq\phi\leq2\pi$, $0\leq\theta\leq\pi$, and $0\leq r\leq R\left(\Delta t\right)$, and using eq. (\ref{MSD1}) yields the number of $A$ particles passing out of $A$ through $\delta S$,
\beq
\Delta{\cal{N}}_A = \frac{5l_A\delta S}{48}\int\limits_0^\infty n_A\left(v,t\right) \sqrt{d\frac{\ln(1+ \phi v\Delta t/l_A)}{\ln{(1/\epsilon)}}} dv \quad ,
\label{flux0}
\eeq
in which $l_A$ is the mean free path in region $A$. The total flux from $A$ to $B$ is then $\Delta{\cal{N}}_T\equiv\Delta{\cal{N}}_A-\Delta{\cal{N}}_B$. To determine its sign analytically is difficult, but using (\ref{flux0}) to solve numerically for the difference between the fluxes establishes that $\Delta{\cal{N}}_T>0$ for any arbitrary density difference, $\delta\rho$. The numerical calculation yields the generic stability diagram shown in Fig. \ref{fig:StabilityPhases}, for $\epsilon = 0.8$, $\rho_1=1/l_0^3$ and $v_1 = 1000 l_0/\Delta t$. It shows that a flow from $A$ to $B$ is possible when the corresponding mean spatial velocity fluctuation, $\delta v = v_A - v_B$, exceeds a threshold value (the red line) and it illustrates two key points.
(i) At small local density differences, vanishingly small velocity fluctuations are required to destabilise the uniform density, establishing that the uniform density is {\it always unstable}.
(ii) The threshold velocity fluctuation increases {\it sub-linearly} with the density difference. Therefore, once particles start escaping from $A$ to $B$, the flux is more sensitive to the reduction in velocities, because of the increased collision rate, than to the counter effect of the density increase. The sharp drop in the speed distribution with time (Fig. \ref{fig:Pv}) overwhelms the slower increase of the other terms in (\ref{flux0}) and this is the main reason for the instability.
Moreover, particles starting at $dV$ with speed $v$ arrive at $\delta S$ after suffering on average $(r/l_0)^2$ collisions, reducing their speeds exponentially to $v\epsilon^{(r/l_0)^2}$. Therefore, $B$ particles arrive at $\delta S$ with much lower speeds, which increases further  the flux from $A$ to $B$ and sharpens the instability.  This phenomenon has been observed in cooling granular gases~\cite{GoZa93,Lu05,Kuetal97,Maetal08} and it leads eventually to formation of low density cosmic regions \cite{GrTh78,Lietal95}. \\
   
\noindent{\bf Cluster growth} 

The above instability densifies small regions until the mean free path gets close to the diameter of the particle scattering cross section and the mean particle speed is very low. These can then be regarded as cluster nuclei. There is a finite probability that such clusters break up when hit by high-momentum particles, a probability that decreases with cluster size. The following analysis focuses on small clusters that survive these teething troubles. 
The cluster grows by low-velocity gas particles colliding with it and are left with insufficient momentum to scatter away. Such momentum loss is used, e.g. in energy dampers \cite{Koetal15}. The following is an effective medium analysis derivation of the cluster growth rates, their dependence on the cluster shape, and determination whether filaments and sheets are more or less likely to emerge than compact clusters. This analysis is valid at longer time scales than in the previous section.

Starting with relatively compact clusters, imagine a $d$-dimensional system of $N$ particles, comprising an isolated small cluster of $N_c (\ll N)$ particles, surrounded by a large gaseous medium. The total system volume is $V$, of which the cluster occupies a volume $V_c\ll V$. 
The number density in the gas, $\rho_g=(N-N_c)/(V-V_c)$ is, by definition, lower than in the nuclear cluster, $\rho_c=N_c/V_c$, which is presumed uniform. 
The velocity distribution in the gas, $n(v)$, is presumed isotropic and position independent. Since $N_c \ll N$ the gas can be regarded as a particle reservoir with $n(v)$ changing negligibly as particles join the cluster.
Simplest is to assume that particles remain close to the cluster only when they collide with it at velocities lower than some threshold $v_c$. Since colliding particles interact only locally with the cluster surface, $v_c$ is assumed independent of the cluster size for clusters larger than a few particles. 
The cluster growth rate is then equal to the number of particles with $v\leq v_c$ colliding with its surface, $S_c$, per unit time:
\begin{equation}
\frac{dN_c}{dt} + \int\limits_0^{\mid v\mid = v_c} n(\vec{v}) \left(\oint_{S_c} \vec{v}\cdot d\vec{S_c}\right)d^3\vec{v} = 0  \ ,
\label{Growth0}
\end{equation}
where $\vec{S_c}$ is normal to the surface and points away from the cluster. For isotropic and uniform velocity distributions, this relation simplifies to
\begin{equation}
\frac{dN_c}{dt} = \rho_g u S_c \ ,
\label{Growth1}
\end{equation}
where $u\equiv\int_0^{v_c} v n(v) dv/(\sqrt{d}\rho_g)$ is the mean normal-to-the-surface velocity of the particles that can join the cluster. 
Substituting in (\ref{Growth1}) for the cluster's surface area, $S_c=C_d\left(N_c/\rho_c\right)^{1-1/d}\sim V_c^{1-1/d}$, and for $\rho_g$, yields
\begin{equation}
\frac{dN_c}{dt} = \frac{N-N_c}{V-V_c}C_d u \left(N_c/\rho_c\right)^{1-1/d} \ .
\label{Growth2}
\end{equation}
Defining the global number density $\rho_0\equiv N/V$ and rearranging, the equation governing the growth rate of the cluster fractional mass, $f\equiv N_c/N$, is
\begin{equation}
\frac{d f}{d t} = C_d u \left(\frac{N}{\rho_c}\right)^{-1/d} \frac{1-f}{\left(\rho_c/\rho_0\right) - f} f^{1-1/d} \ .
\label{ClusterGrowth}
\end{equation}
Noting that $\beta\equiv \rho_c/\rho_0 > 1$, integrating, and choosing the initial condition $f(t=0)=0$, one obtains 
\begin{eqnarray}
\frac{1}{\cal{L}} & \int & u dt = \nonumber \\
& = & \left\{1 + \beta \left[{}_2 F_1(1,1/d;1+1/d;f) - 1\right]\right\}f^{1/d} \ ,
\label{ClusterGrowth1}
\end{eqnarray}
with ${\cal{L}}\equiv \frac{1}{C_d} \left(\frac{N}{\rho_c}\right)^{1/d}$ a constant length of order of the system linear size and ${}_2 F_1$ the Gauss hypergeometric function. Relation (\ref{ClusterGrowth1}) holds for any velocity distribution $n(v)$ and, in particular, for the one derived in (\ref{PDFv2}).  
However, by the above effective medium assumption, $n(v)$ hardly changes on the time scale of the initial growth, and therefore $u$ is practically constant. Then the left hand side of (\ref{ClusterGrowth1}) reduces to $t/\tau_c$, where $\tau_c={\cal{L}}/u$ is roughly the time it takes a particle with speed $u$ to cross the entire system uncollided. 

Eq. (\ref{ClusterGrowth1}) describes the growth of a compact cluster, but is the compact shape growing stably? 
As particles collide with the cluster they transfer much of their kinetic energy to it. This increases the kinetic energy of the cluster particles, which cannot be fully dissipated via the restitution-based mechanism. Yet, simulations show that clusters do not appear to `heat up' compared to the surrounding gas~\cite{MiLu04,Lu05,Goetal14}. The main reason is that, although the collisional momentum dissipates exponentially with distance away from the collision point \cite{Poetal05,Koetal15}, it may suffice to eject particles from the surface opposite to the collision. Such particles would be ejected if the momentum transferred to it exceeds a threshold $v_e$. Another source of energy absorption could be the excitation of intra-particle degrees of freedom, but this is negligible compared with particle ejection.

\begin{figure}[htbp]
      \centering
      \includegraphics[width=.4\textwidth]{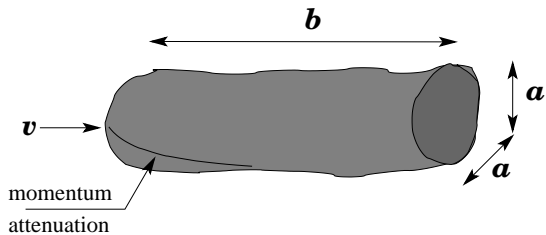}
      \caption{An elongated cluster of dimensions about $a\times a\times b$. The momentum of a colliding particle, illustrated by the thick line, attenuates exponentially away from the collision point.} 
      \label{LongClusterGrowth}
   \end{figure}

To analyse a non-spherical cluster, let us model clusters, such as sketched in Fig. \ref{LongClusterGrowth}, as a rectangular prism of $a\times a\times b$.  If the in-cluster momentum dissipation rate of a colliding particle is $e^{-r/l_p}$ and the opposite face is a distance $L$ ($=a$ or $b$) away then the momentum at the opposite surface is reduced by $e^{-L/l_p}$ and particles would be ejected if the collision velocity, $v$, satisfies $ve^{-L/l_p}\geq v_e$. 
   
The exact decay length, $l_p$, is difficult to determine and, for $\epsilon$ very close to unity, can be quite large compared with the particle size. Assuming, for simplicity, that one collision ejects only one particle, the velocity of the ejected particle is $ve^{-L/l_p}$. The number of aggregating particles per unit surface area per unit time is $\int_0^{v_c} vn(v)dv/(6\rho_g)$, and the number of ejected particles from the opposite end per unit area per unit time is $\int_{v_ee^{L/l_p}}^{v_{max}} vn(v)dv/(6\rho_g)$, where $v_{max}$ is the highest speed in the gas. The cluster's sides grow  at different rates:
\begin{eqnarray}
\frac{\rho_c dL}{dt} & = & 2\int\limits_0^{v_c} vn(v)dv - \frac{2}{6}\int\limits_{v_ee^{L/l_p}}^{v_{max}} vn(v)dv  \ .
\label{eqgrowth}
\end{eqnarray}
Using (\ref{eqgrowth}) for $L=a$ and $b$ and dividing, yields the relative growth rates:
\begin{equation}
\frac{db/dt}{da/dt} = 1 + \frac{\frac{1}{6}\int\limits_{v_ee^{a/p}}^{v_ee^{b/p}} vn(v)dv}
{\int\limits_0^{v_c} vn(v)dv - \frac{1}{6}\int\limits_{v_ee^{a/p}}^{v_{max}} vn(v)dv} \ .
\label{relgrowth0}
\end{equation}
Inspecting (\ref{relgrowth0}), $db/dt > da/dt$ when $b>a$ and vice versa, i.e. the longer side grows faster than the shorter one. Thus, the compact cluster solution is {\it unstable} - clusters grow either as long filaments or thin sheets. 
This instability is independent of the specific form of $n(v)$, which determines only the value of $(db/dt - da/dt)$, namely, the strength of the instability. 

When $b\gg a$, we can neglect the growth of $a$ and use eq. (\ref{Growth1}) with $S_c\approx 4ab$ and $N_c = ba^2\rho_c$. Following the steps leading from eq. (\ref{Growth0}) to eq. (\ref{Growth2}), we have
\begin{equation}
\frac{db}{dt} = \frac{4u}{a\rho_c}\frac{N-N_c}{V-V_c}b \ .
\label{relgrowth1}
\end{equation}
Defining $\rho_0\equiv N/V$, $\alpha\equiv\rho_c/\rho_0$, and $f\equiv N_c/N$, this equation can be rewritten as 
\begin{equation}
\frac{df}{dt} = \frac{4u}{a}\frac{\left(1 - f\right)f}{\alpha - f} \ .
\label{Growth5}
\end{equation}
In principle, $\alpha$ increases with time as gas particles are lost to the cluster and the universe expands. The former is negligible for $N_c\ll N$. 
By setting $\alpha$=constant, for which (\ref{Growth5}) can be solved explicitly, we obtain an {\it upper bound} on the growth rate:
\begin{equation}
e^{t/\theta} = \frac{f}{f_0}\left(\frac{1-f_0}{1-f}\right)^{1-1/\alpha} \ ,
\label{relgrowth2}
\end{equation}
with  $f_0=N_c(t=0)/N$ and $\theta\equiv a/u$ the time it takes a particle with mean speed $u$  to cross the narrow side of the cluster.
For illustration, this upper-bound solution is plotted in Fig. \ref{ClusterGrowth} together with the numerical solution of (\ref{Growth5}) for an expanding flat universe for $\alpha=1.5$ and $100$. The two become indistinguishable as $\alpha$ increases. As estimated below, $\alpha_0\gg1$ by many orders of magnitude already at recombination, making this a very tight bound. Moreover, since $\alpha-f\approx\alpha$, (\ref{relgrowth2}) simplifies and the elongation rate of filamentary clusters is to very good accuracy,

\begin{equation}
f = \frac{1}{1 + \frac{1-f_0}{f_0}e^{-4t/(\alpha\theta)}} \ .
\label{relgrowth3}
\end{equation}
\begin{figure}[htbp]
  \centering
  \subfigure[]{
 \includegraphics[width=.4\textwidth]{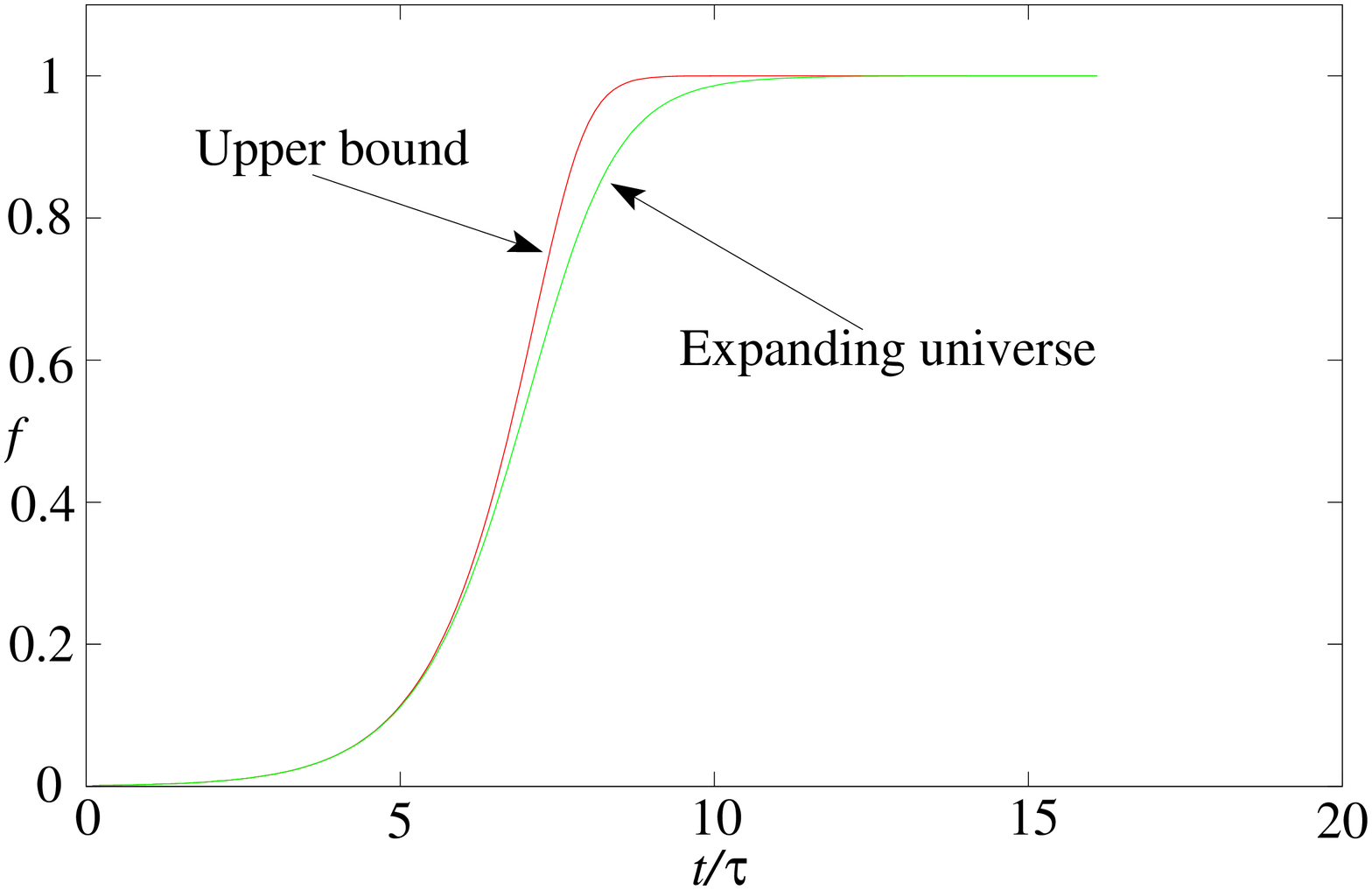}
 \label{ClusterGrowth1p5}
      }
\subfigure[]{
 \includegraphics[width=.4\textwidth]{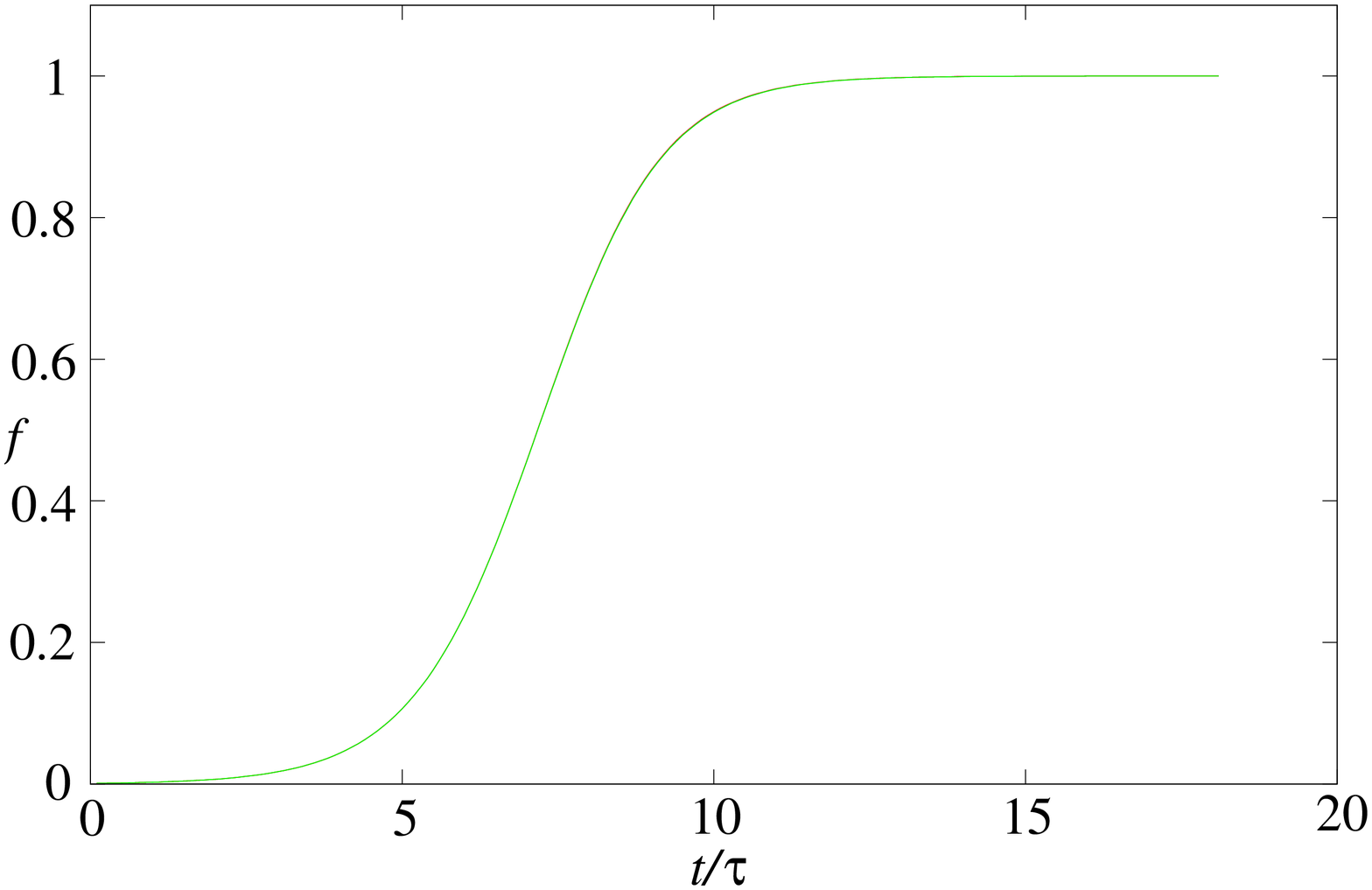}
 \label{ClusterGrowth100p}
    }
      \caption{The growth rate of filamentary clusters in the expanding universe and the upper bound on it for density ratios (a) $\alpha=1.5$ and (b) $100$. Time is measured in units of $\tau = a\alpha_0/(4u)$ and here $a/4u=10^{-2}$. For values of $\alpha>100$ the two are indistinguishable, which is the case after recombination, $\alpha= O\left(10^{18}\right)$.  }
      \label{ClusterGrowth}
   \end{figure}%
   
\noindent{\bf Numerical estimates} 

To check the predictions of (\ref{relgrowth3}), we need to estimate $\alpha$ and $\theta$. 
Approximating the present number density of all matter as about $5$ hydrogen atoms per $1$m$^3$, of which about $4\%$ is baryonic matter and $21\%$ dark matter, gives a present matter number density of about $1.25$ hydrogen atoms per $1$m$^3$. Assuming negligible interaction between mass and radiation since recombination and using the recombination redshift, $z_{*}\approx1100$, yields $\rho_{matter}=\rho_0\approx 1.25\times z_{*}^3\approx1.66\times10^9$m$^{-3}$. 
The intra-cluster mean free path is 2-5 times the effective diameter of a hydrogen atom, $\approx 5\times10^{-10}$m, equivalent to cluster number density $\rho_c\approx 8\times10^{27}$m$^{-3}$. Thence, at recombination, $\alpha_0\equiv\rho_c/\rho_0\approx 4.8\times10^{18}\gg1$.
For a rough estimate of $\theta$, assume thermodynamic equilibrium at temperature $\approx3000$K at $z_{*}$. Using the equipartition principle, the mean speed of the hydrogen atoms was then $\approx8.6\times10^3$m/sec and, taking the value of $u$ as one order of magnitude smaller, say $\sim10^3$m/sec, the value of $\theta$ for a cylindrical cluster of diameter $10^x$m is $\approx10^{x-3}$sec. This yields  
$\alpha_0\theta/4\approx1.2\times10^{x+15}$sec$=3.81\times10^{x+8}$yrs. Expecting $x<0$ at recombination, $\alpha_0\theta/4 < 3.81\times10^{8}$yrs. 

However, it should be noted that several competing mechanisms cause $\alpha$ and $\theta$ to evolve slowly. 
Firstly, $\alpha$ increases as the gas is depleted and the universe expands, which slows down the clustering process. 
Secondly, the decreasing velocities in the gas increases the number of slow particles that can join the cluster. This would accelerate clustering but for the reduction in the flux of the particles onto the cluster's surface, which decelerates clustering. 
Thirdly, the short side of the cluster also grows, albeit more slowly, which increases $\theta$. 
These effects comprise small perturbations on solution (\ref{relgrowth3}), but they accumulate over time. 
Taking all those into consideration needs to be done numerically and this will be reported at a later stage.  \\

\noindent{\bf Conclusion and discussion} 

To conclude, assuming a flat universe, collisional momentum loss has been shown to destabilise the initial cosmic uniform density into formation of clusters. 
Starting from a uniform gas state of relatively cold matter, a self-consistent gravity-free model has been developed for the structural organisation of the cosmic web. 
The inelastic collisions give rise to sub-anomalous particle diffusion, with the MSD increasing {\it logarithmically} with time. 
This result was used to derive the evolution of the velocity distribution in the gas, which turns out to have an unusual form - eq. (\ref{PDFv2}). 
These results were shown to lead to an instability of the uniform matter density, with {\it matter flowing into denser regions}, eventually leading to cluster nucleation. 
The clusters grow by `capturing' colliding particles with insufficient momenta to escape. 
An effective-medium equation has been derived then for the growth rate of compact clusters and solved analytically. However, this solution was shown to be unstable to cluster elongation. The growth rate of clusters into filaments has been derived then explicitly.

The main implication of this model is that a cosmic web of filaments and sheets could have well formed before gravity took effect, which means that gravitational collapse need not be the main mechanism giving rise to it. 
During this era, low-density regions grew as matter accreted into clusters, with the process slowing down as supply of particles dwindled and depleted voids appeared. 
When this mechanism ran its course, gravity was left as the dominant mechanism and the massive non-compact clusters started collapsing into more compact forms. This agrees with the idea that the web is much older than the compact formations. Intriguingly, this may suggest that the current compact formations could have started as locally anisotropic clusters, a conclusion that may have observable consequences. In turn, such observations could provide a test of this model. 
This first-principles approach circumvents the need to invoke conventional hydrodynamic equations, whose validity for dense particulate gases is far from clear. 

This analysis has limitations. One is ignoring the universe's expansion in the calculation of the diffusion processes. The expansion slows down even further the velocities between collisions. However, at the recombination's high densities this is expected to be a perturbation that may change the conclusions reached here quantitatively somewhat but not at all qualitatively. Including the expansion's effect can be done by adding a small drag, but this would necessitate a numerical solution of the equations and obscure the clarity of the analytical treatment. The expansion also affects the rate of cluster growth by both slowing the gas velocities and reducing the gas density. These are also expected to be a perturbation, especially since the two have opposite effects on clusters growth, as discussed in the previous section. 
Another limitation arises from the effective medium modelling of the cluster growth, which is valid as long as clusters are sufficiently far apart to neither interact gravitationally nor compete over the surrounding gas particles. Such a competition eventually becomes unavoidable as clusters grow and the gas rarefies. Fast-growing clusters suppress the growth of other clusters by depleting their neighbourhoods. This `rich-get-richer' dynamics leads directly to an abundance of elongated formations surrounding by large voids.

An immediate question arises: the model explains how baryonic matter organises into a web-like structure, but not how the cosmic web contains an abundance of dark matter. There are two possible answers. One is that the clusters of baryonic matter could have acted as cluster nuclei and dark matter particles aggregated around them gravitationally. 
Another, more speculative, is that dark matter particles may also scatter off one another, however slightly, in which case the model could apply to dark matter as well and the process could have started even before recombination, with hydrogen atoms joining the dynamics later. 

Finally, the bouncing barrier problem~\cite{BouncBar}, posed by eq. (\ref{Gravity}), also hinders modelling planetesimals formation from mm-size particles. It is possible that a similar mechanism led to dust clustering in massive filaments and sheets, after which gravity took hold and compressed the clusters into more compact shapes, such as those observed today. It would be interesting to test this possible scenario by searching for traces of the initial cluster anisotropies in the current compact planets. \\

\noindent{\bf Acknowledgments}

RB is grateful to Dr W. Handley for useful suggestions and critical reading, to Prof. M.-Y. Hou for discussions, and to S. Blumenfeld for help with Python and graphics. RB acknowledges the hospitality of the Cavendish Laboratory.

\end{document}